\documentclass[
 preprint,
 amsmath,amssymb,
 aps,
prb,
]{revtex4-1}

\usepackage{graphicx}
\usepackage{dcolumn}
\usepackage{bm}
\usepackage{hyperref}
\usepackage[mathlines]{lineno}

\begin{document}

\preprint{APS/123-QED}

\title{Doping-induced persistent spin helix with large spin splitting in monolayer SnSe}

\author{Moh. Adhib Ulil Absor}
\email{adib@ugm.ac.id} 
\affiliation{Department of Physics, Universitas Gadjah Mada BLS 21 Yogyakarta Indonesia.}%

\author{Fumiyuki Ishii}%
\affiliation{Faculty of Mathematics and Physics Institute of Science and Engineering Kanazawa University 920-1192 Kanazawa Japan.}%

\date{\today}

\begin{abstract}
Finding a new class of materials supporting a long spin lifetime is essential in development of energy-saving spintronics, which is achievable by using a persistent spin helix (PSH) materials. However, for spintronic devices, the PSH states with large spin splitting is required for operation at room temperature. By employing first-principles calculations, we show that the PSH states with large spin splitting are achieved in the SnSe monolayer (ML) functionalized by a substitutional halogen impurity. We find the PSH states in the Fermi level where $k$-space Fermi surface is characterized by the shifted two loops, dominated by out-of-plane spin orientations. We clarify the PSH states in term of an effective $\vec{k}\cdot \vec{p}$ Hamiltonian obtained from symmetry consideration. Finally, large spin-orbit strength in the PSH states with a substantially small wavelength are found, rendering that this system is promising for the development of an efficient and high-density scalable spintronic devices operating at room temperatures.     
\end{abstract}

\pacs{Valid PACS appear here}
\maketitle

\section{INTRODUCTION}
Recent development of spintronics relies on the new pathway for exploiting carrier spins in semiconductors by utilizing the effect of spin-orbit coupling (SOC)\cite{Manchon}. When the SOC occurs in a system with sufficiently low crystalline symmetry, an effective magnetic field or known as a spin-orbit field (SOF) $\vec{\Omega}(k)\propto [\vec{E}\times \vec{p}]$ is induced, where $\vec{E}$ is the electric field originated from the crystal inversion asymmetry and $\vec{p}$ is the momentum, that leads to spin splitting energy \cite {Rashba,Dresselhauss}. However, due to momentum-dependent of the SOF, electron scatterings randomize the spins, which induces the fast spin decoherence through the Dyakonov-Perel mechanism \cite {Dyakonov}. Accordingly, spin lifetime reduces, which limits the performance of potential spintronic devices, e.g., the spin field effect transistor (SFET)\cite {Datta}. To overcome this problem, finding a novel structure supporting an extended spin lifetime is an important task, which is achievable by using persistent spin helix (PSH) materials\cite{Bernevig,Schliemann,Passmaan,Koralek,Walser,Kohda,Sasaki,Absor1,Yamaguchi,LLTao}.

The PSH is established on a system having unidirectional SOF, satisfying SU(2) symmetry \cite{Bernevig}. Here, a uniform spin textures in the momentum $k$-space is induced, resulting in a robust against spin-independent scattering. Such a PSH condition is achieved, in particular, if the strength of the Rashba and Dresselhauss SOC is equal, as recenly reported experimentally on various (001)-oriented QW such as GaAs/AlGaAs QW \cite {Koralek,Walser,Schonhuber} and InGaAs/InAlAs QW \cite {Ishihara,Kohda,Sasaki}. On the other hand, the PSH state can also be achieved on the (110)-oriented QW where the SOC is purely characterized by the Dreseelhauss effect \cite{Bernevig}. Here, the SOF is enforced to be parallel with the spins in the unidirectional out-of-plane orientation, which has been experimentally observed on (110)-oriented GaAs/AlGaAs QW \cite{Chen}. This PSH state was also predicted for a wurtzite ZnO (10-10) surface \cite {Absor1} and strained LaAlO3/SrTiO3 (001) interface \cite {Yamaguchi}. Although the PSH has been experimentally realized in various QW systems \cite {Ishihara,Kohda,Sasaki,Koralek,Walser,Schonhuber}, it is practically non-trivial. This is due to the fact that achieving the PSH state requires to control precisely the strength of the Rashba and Dresselhauss SOC, for instant, by tuning the QW width \cite{Koralek,Walser} or using an external electric field \cite{Ishihara,Kohda,Sasaki}. 

Since the orientation of the SOF $\vec{\Omega}(k)$ is enforced by the direction of the electric field $\vec{E}$, the PSH states may appear intrinsically on a material exhibiting internal electric polarization. Here, two-dimensional (2D) ferroelectric materials such as tin monoselenide (SnSe) monolayer (ML) comprises promising candidate for the PSH material since it was predicted to have large in-plane electric polarization \cite{Fei,Lopez}. This electric polarization is expected to induce electric field $\vec{E}$ in the in-plane direction, which imposes the SOF becoming out-of-plane unidirectional \cite {Absor1}. Recently, the SnSe ML has been succsesfully synthesized experimentally using vapor deposition processes \cite{Zh} or the one-pot route \cite{Li}, which shows high carrier mobility \cite{Xu} and anisotropic electrical conductivity \cite{Shi}. Therefore, the experimental realization of the SnSe ML as a PSH material is plausible. Although the SnSe ML possibly supports the PSH states, this material posses small spin splitting energy ($\approx 50$ meV) \cite{Shi,Gomes_a}, which provides a disadvantage for spintronic devices. Therefore, finding a suitable adjustment method to achieved the large spin splitting in the SnSe ML is a great challenge. 

In this work, we proposed to use a halogen doping for inducing large spin splitting in the SnSe ML. Introducing the halogen doping is expected to enhance the in-plane electric polarization through in-plane ferroelectric distortion, which evidently gives rise to the large spin-splitting. In fact, the large spin splitting induced by a halogen doping has been recently reported on various 2D transition metal dichalcogenides (TMDs) such as WSe$_{2}$ \cite{Guo_S} and PtSe$_{2}$ MLs \cite{Absor3}. By employing fully relativistic density-functional theory (DFT) calculations, we find that the PSH state with large spin splitting is achieved on the SnSe ML functionalized by a halogen substitutional doping. We identify the PSH states at Fermi level where the the $k$-space Fermi surface is characterized by the shifted two loops, dominated by the out-of-plane spin textures. We analysed the PSH states in term of an effective $\vec{k}\cdot \vec{p}$ Hamiltonian obtained from symmetry consideration. Finally, we estimate the spin splitting-related parameters such as spin-orbit strength and the wavelength of the PSH states, and discuss the possible application in the  spintronic devices.   

\section{Computational Details}

We performed first-principles electronic structure calculations based on the non-collinear density functional theory (DFT) within the generalized gradient approximation (GGA) \cite {Perdew} using the OpenMX code \cite{Openmx}. We used norm-conserving pseudopotentials \cite {Troullier}, and the wave functions are expanded by the linear combination of multiple pseudoatomic orbitals generated using a confinement scheme \cite{Ozaki,Ozakikino}. In our DFT calculations, the SOC is included by using $j$-dependent pseudopotentials \citep{Theurich}. The basis functions of each atoms are two $s$-, two $p$-, two $d$-character numerical pseudo-atomic orbitals. 

\begin{figure}
	\centering
		\includegraphics[width=1.0\textwidth]{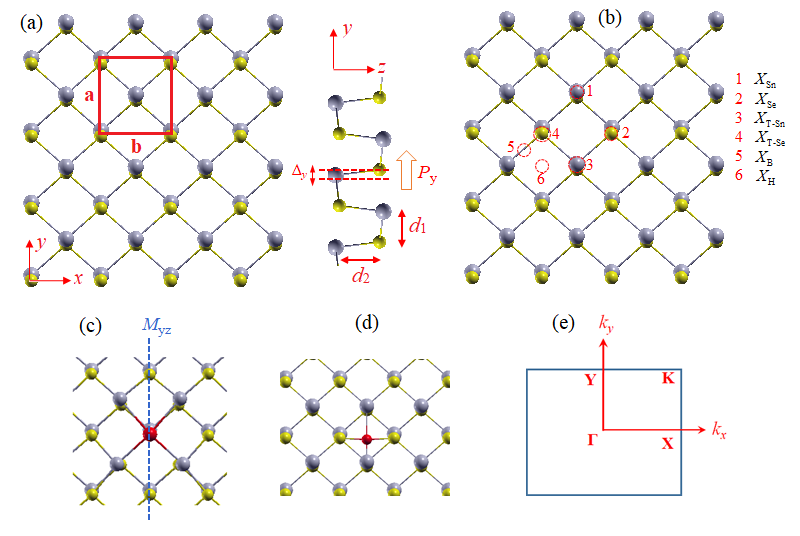}
	\caption{ (a) Structure of the pristine SnSe monolayer in the top view and side views is shown. The unit cell is indicated by the rectangle with red line, which is characterized by $a$ and $b$ lattice parameters. The layers sit on the $x-y$ plane with the in-plane distortion $\Delta_{y}$ along the $y$-axis. The bondlength between Sn and Se in the in-plane ($d_{1}$) and out-of-plane ($d_{2}$) directions are indicated. $\Delta_{y}$ represents in-plane distortion along the $y$-direction. (d) Various possible doping configurations such as (1) $X_{\text{Sn}}$, (2) $X_{\text{Se}}$, (3) $X_{\text{T-Sn}}$, (4) $X_{\text{T-Se}}$, (5) $X_{\text{B}}$, and (6) $X_{\text{H}}$, are shownn. The optimized structures for (c) I$_{\text{Se}}$ and (d) I$_{\text{H}}$ as a representative example of the $X_{\text{Se}}$ and $X_{\text{H}}$, respectively. The red ball indicates the I atom. Mirror symmetry $M_{yz}$ operation in the optimized I$_{\text{Se}}$ is indicated by the blue-dashed lines. (e) First Brillouin zone of the SnSe ML is shown, where the high symmetry points such as $\Gamma$, $X$, $Y$, and $K$ points are indicated.}
	\label{figure:Figure1}
\end{figure}

We used a periodic slab to model the SnSe ML, where a sufficiently large vacuum layer (20 \AA) is applied to avoid interaction between adjacent layers. The crystal structures of the SnSe ML with ferroelectric phase \citep{Fei} were used. Here, we used the axes system where layers are chosen to sit on the $x-y$ plane, i.e., perpendicular to the $z$ direction [Fig. 1(a)]. The $x$ axis is taken to be parallel to the puckering direction, which is similar to the system used for black phosphorus \cite{Popovic}. The calculated lattice parameters are $a=4.36$ \AA\ and $b=4.30$ \AA, which is in a good agreement with previous calculation \cite {Gomes1,Gomes,Xu,Haleoot}.

We then introduced a doping with external element taken from the halogen (Cl, Br, and I) atoms. Here, the halogen dopant can either occupy a substitutional position or adsorb on the surface following Boltzmann distribution to form a point defect. In the case of the substitutional doping, we considered $X_{\text{Sn}}$ and $X_{\text{Se}}$ defined as the halogen $X$ substitution on the Sn and Se sites, respectively. On the other hand, for the case of the absorption, various halogen doping positions such as top of the Sn atom ($X_{\text{T-Sn}}$), top of the Se atom ($X_{\text{T-Se}}$), in-plane bridge between the Sn and Se atoms ($X_{\text{B}}$), and hollow surface ($X_{\text{H}}$) were considered [Fig. 1(b)]. To model these doping systems, we construct a 4x4x1 supercell of the SnSe ML with 64 atoms. The larger supercells (5x5x1 and 6x6x1 supercells) are used to test our calculation results, and we confirmed that it does not affect the main conclusion. The geometries were fully relaxed until the force acting on each atom was less than 1 meV/\AA. 

To confirmed stability of the doped SnSe, we calculated the formation energy $E_{f}$ defined as 
\begin{equation}
\label{1}
E_{f}=E_{\text{SnSe}:X}-E_{\text{SnSe}}+\sum_{i}n_{i}(E_{i}+mu_{i}).
\end{equation}  
where $E_{\text{SnSe}:X}$ and $E_{\text{SnSe}}$ are the total energy of the doped and pristine SnSe ML, respectively. In Eq. (\ref{1}), $\mu_{i}$ is the chemical potentials of constituent $i$, while $E_{i}$ is the total energy calculated for the most stable element crystal. $n_{i}$ is the number of the component atoms $i$ added to (negative $n_{i}$), or taken from (positif $n_{i}$) the host. There are many thermodynamic limits on the achievable values of the chemical potential under equilibrium growth conditions \cite {Wei_1}. first, to avoid precipitation of the elemental dopant $X$ and the host elements (Sn and Se), $\mu_{i}$ are bound by
\begin{equation}
\label{2}
\mu_{X}\leqslant 0, \mu_{\text{Sn}}\leqslant 0,\mu_{\text{Se}}\leqslant 0.
\end{equation} 
Second, the sum of the chemical potentials of all component elements should be equal to the host compound:
\begin{equation}
\label{3}
\mu_{\text{Sn}}+\mu_{\text{Se}}=E_{f(\text{SnSe})}
\end{equation}
where $E_{f(\text{SnSe})}$ is the formation energy of the SnSe ML. In our calculation, we found that $E_{f(\text{SnSe})}=-0.87$ eV, which is in a good agreement with previous result obtained by Tang et. al. \citep{Tang}. Finally, to avoid the formation of secondary phases between the halogen dopants and host element, $\mu_{i}$ is limited by:
\begin{equation}
\label{4}
\mu_{\text{Sn}}+2\mu_{\text{Se}}<E_{f(\text{SnSe}_{2})}
\end{equation}
\begin{equation}
\label{5}
\mu_{\text{Sn}}+2\mu_{X}<E_{f(\text{Sn}X_{2})}
\end{equation}
where $E_{f(\text{SnSe}_{2})}$ and $E_{f(\text{Sn}X_{2})}$ are the formation energy of SnSe$_{2}$ and Sn$X_{2}$ MLs.

\begin{table}[ht!]
\caption{The calculated formation energy $E_{f}$ (in eV) of the halogen-doped SnSe ML under both the Sn- and Se-rich conditions. Here, several halogen dopants such as Cl, Br, and I atoms are considered.} 
\centering 
\begin{tabular}{c  c  c  c  c  c  c  c  c  c} 
\hline\hline 
    &  & \multicolumn{2}{c}{Cl}  &  & \multicolumn{2}{c}{Br}  &   &  \multicolumn{2}{c}{I} \\ [0.5ex]
     \cline{3-4} 
     \cline{6-7} 
     \cline{9-10}   
Impurity type &  & Sn-rich (eV)  & Se-rich (eV) &  & Sn-rich (eV)  & Se-rich (eV) &  & Sn-rich (eV)  & Se-rich (eV) \\[0.5ex] 
\hline 
Substitution &  &   &  & &   &  & &   &   \\
$X_{\text{Sn}}$  & & 2.12 & 1.57 & & 2.47  & 1.8 & & 0.2.74  & 1.91  \\
$X_{\text{Se}}$ &  & 0.76 & 0.42 & & 0.81  & 0.58 & & 0.94  & 0.77  \\
Absorption  & &   &  & &   &  & &   &   \\
$X_{\text{T-Sn}}$ & & 0.85 & 1.10 & & 0.92  & 1.23 & & 1.09  & 1.34  \\
$X_{\text{T-Se}}$ &  & 1.2 & 2.01 & & 1.55  & 2.43 & & 1.91  & 2.78  \\
$X_{\text{B}}$ &  & 1.63 & 2.21 & & 1.95  & 2.79 & & 2.19  & 2.98  \\
$X_{\text{H}}$ & & 0.78 & 0.91 & & 0.82  & 1.04 & & 0.98  & 1.11  \\
[1ex] 
\hline\hline 
\end{tabular}
\label{table:Table} 
\end{table}

\section{RESULT AND DISCUSSION}

First, we briefly discuss the structural and electronic properties of the pristine SnSe ML. From symmetry point of view, crystal structure of the SnSe ML is non-centrosymmetric, which is characterized by three non-trivial symmetry operations, namely a vertical mirror plane ($M_{yz}$) parallel to the $y-z$ plane, a two-fold screw rotation along the axes parallel to the $x$ direction, and a glide reflection on a plane parallel to the $x-y$ plane. Here, the Sn and Se atoms have an opposing displacement within the $y-z$ plane [Fig. 1(a)], which induces in-plane ferroelectric distortion $\Delta_{y}$ in the $y$-direction [Fig. 1(b)]. This distortion will generate in-plane electric polarization $P_{y}$ \cite{Fei,Lopez}, and then induces in-plane electric field $E_{y}$, which plays an important role for inducing the PSH states.

\begin{figure}
	\centering
		\includegraphics[width=0.85\textwidth]{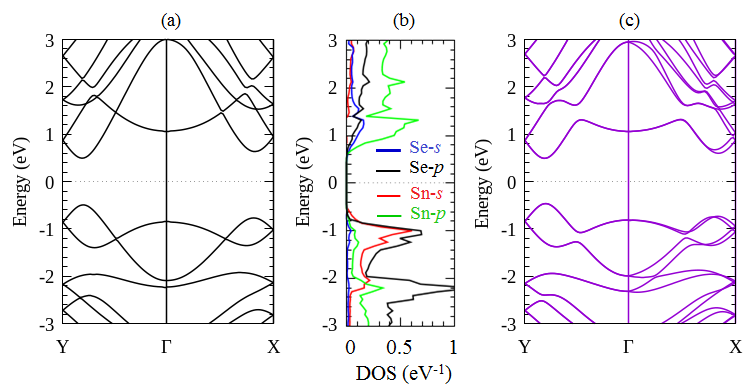}
	\caption{Electronic band structures and density of states (DOS) of the pristine SnSe ML. (a) The band structures calculated without the SOC corresponding to the DOS projected to the atom (b). The band structures calculated with the SOC are shown.}
	\label{figure:Figure2}
\end{figure}

Figure 2(a) shows the calculated result of the electronic band structures of (1x1) unit cell of the SnSe ML. The first Brillouin zone used in our calculations is shown in Fig. 1(e). Consistent with previous result \cite{Shi,Gomes_a}, we find two equivalent band dispersions near the gap, which is located along the $\Gamma-X$ and $\Gamma-Y$ lines [Fig. 2(a)]. However, both the valence band maximum (VBM) and conduction band minimum (CBM) are located on the same point in the $\Gamma-Y$ line, resulting in the direct gap semiconductor with energy gap of 0.95 eV. Our calculated density of states (DOS) projected to the atom confirmed that the Sn-$p$ and Se-$s$ orbitals contribute mostly to the VBM, while the CBM is mainly originated from the contribution of the Sn-$p$ and Se-$s$ orbitals [Fig.2(b)]. 

Turning the SOC gives rise to the spin-split bands in all the Brillouin zone except for the $\Gamma-Y$ line [Fig. 2(c)]. The observed spin degeneracy along the $\Gamma-Y$ line is due to the fact that this line is parallel to the direction of the in-plane electric field $E_{y}$ in which the $M_{yz}$ mirror symmetry is preserved. Away from the spin-degenerate $\Gamma-Y$ line, the spn splitting is established. Especially in the CBM along $\Gamma-X$ line, the spin splitting energy of 51.6 meV is observed, which is in a good agreement with previous report ($\approx 50$ meV) \cite{Shi,Gomes_a}. Because the spin splitting is suppressed by the in-plane electric field $E_{y}$, it is expected that the spin directions at each spin-split bands are enforced to be perpendicular (out-of-plane) to the SnSe ML surface, which induces the PSH states. This mechanism for inducing the PSH states is similar to the recently reported PSH states on the (110)-oriented GaAs/AlGaAs QW \cite{Chen} and ZnO (10-10) surface \cite {Absor1}.

Next, we consider the effect of the halogen doping on the structural and electronic properties of the SnSe ML. We confirmed the stability of the doped systems ($X_{\text{Sn}}$, $X_{\text{Se}}$, $X_{\text{T-Sn}}$, $X_{\text{T-Se}}$, $X_{\text{B}}$, and $X_{\text{H}}$) by the calculated formation energy $E_{f}$ shown in Table 1.  We find that the $X_{\text{Se}}$ and $X_{\text{H}}$ have the lowest $E_{f}$ under Se-rich and Sn-rich conditions, respectively, indicating that both the systems are the most stable halogen doping formed in the SnSe ML. This is consistent with earlier report that the halogen atom can be easily formed in the group-IV monochalcogenide through substitution on the chalcogen site as found in the bulk SnSe \cite{Zhou}, bulk SnS \cite{Malone}, and GeSe ML \citep{Ao}, or by absorption on the hollow site of the surface as found in the GeSe ML \cite{Zeyu}. On the contrary, the formation of the other dopings ($X_{\text{Sn}}$, $X_{\text{T-Sn}}$, $X_{\text{T-Se}}$, and $X_{\text{B}}$) is highly unfavorable due to the required electron energy. This is due to the fact that the present of the halogen doping induces covalent bonding between the Sn and the halogen $X$ atoms, which is stabilized by breaking the crystal symmetry, thus increasing the $E_{f}$.      

\begin{table}[ht!]
\caption{Structural-related parameters in the pristine and doped ($X_{\text{Se}}$, $X_{\text{H}}$) systems. In the case of the $X_{\text{Se}}$, the bondlength (in \AA) between the Sn and halogen $X$ or Se atoms in the in-plane ($d_{1}$) and out-of-plane ($d_{2}$) directions, and the in-plane distortion $\Delta_{y}$ (in \AA) along the $y$-direction, are shown. For the case of the $X_{\text{H}}$, the average bondlength (in \AA) between the Sn or Se and halogen $X$ atoms ($\bar{d}_{\text{Sn-X}}$, $\bar{d}_{\text{Se-X}}$) are also given. } 
\centering 
\begin{tabular}{c c c c c c c c c c c } 
\hline\hline 
  System &  &   $d_{1}$ (\AA)&  &  $d_{2}$ (\AA)& & $\Delta_{y}$ (\AA) &  &   $\bar{d}_{\text{Sn-X}}$ (\AA) &  & $\bar{d}_{\text{Se-X}}$ (\AA)  \\ 
\hline 
Pristine   &  &  2.89 & & 2.76 &  &    0.34 &  & - &   &  -    \\
Substitutional doping $X_{\text{Se}}$ & & & & & & & & & \\                                                       
Cl$_{\text{Se}}$  & & 3.20 &  & 2.80 &  &  0.39  &  & - &   &  -   \\ 
Br$_{\text{Se}}$   & &  3.24 & &    2.84 &  &     0.42  &  &   - &  &   -    \\
I$_{\text{Se}}$    &  &   3.37 &  &    3.09 & &     0.44 &   &  - &  &  -   \\
Absorption doping $X_{\text{H}}$ & & & & & & & & & \\
Cl$_{\text{H}}$  & & - &  & - &  &  -  &  &  3.39&   & 4.14    \\ 
Br$_{\text{H}}$   & & -  & &  -   &  &   -    &  &  3.42  &  & 4.21      \\
I$_{\text{H}}$    &  & -  &  &   -  & &  -    &   & 3.50  &  &  4.28   \\
\hline\hline 
\end{tabular}
\label{table:Table 1} 
\end{table}

\begin{figure}
	\centering
		\includegraphics[width=1.0\textwidth]{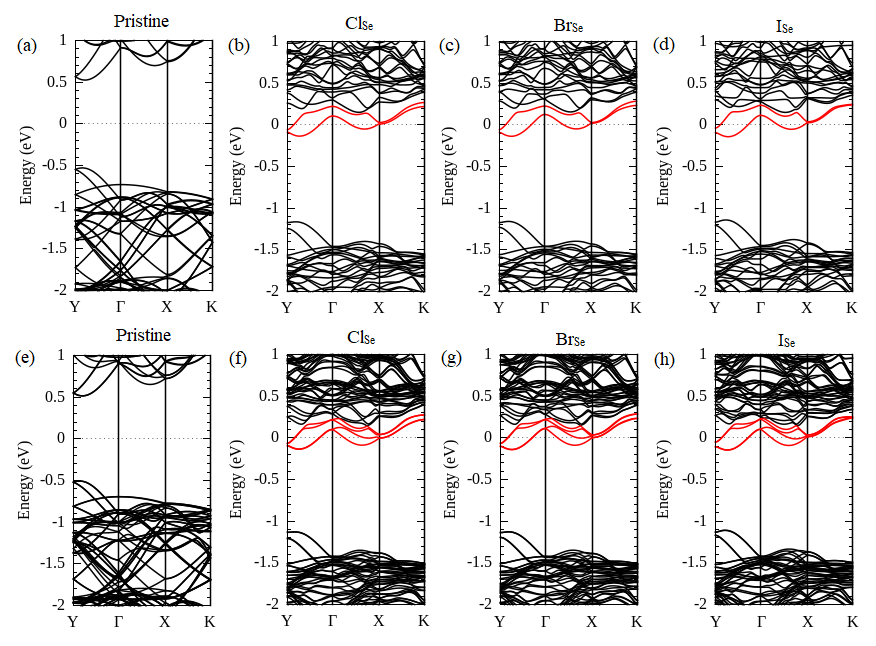}
	\caption{Electronic band structures of $X_{\text{Se}}$ and the pristine systems. The calculated band structures without the SOC for: (a) the pristine, (b) Cl$_{\text{Se}}$, (c) Br$_{\text{Se}}$, and (d) I$_{\text{Se}}$, and with the SOC for: (e) the pristine, (f) Cl$_{\text{Se}}$, (g) Br$_{\text{Se}}$, and (h) I$_{\text{Se}}$. The dashed black lines show the Fermi level. The red lines indicate the localized impurity bands (LIB) around the Fermi level.}
	\label{figure:Figure3}
\end{figure}

When the doping is introduced, the relaxation substantially changes the pristine atomic positions. Here, we consider both the $X_{\text{Se}}$ and $X_{\text{H}}$ because they have the lowest $E_{f}$ among the other doping systems. In the case of the $X_{\text{Se}}$, the relaxation suppressed that only mirror symmetry $M_{yz}$ still remains [Fig. 1(c). Since the bondlength between the Sn and halogen $X$ atoms in the in-plane ($d_{1}$) and out-of-plane ($d_{2}$) directions  are much larger than that between the Sn and Se atoms in the pristine system [see Table 2], a significant in-plane distortion $\Delta_{y}$ is induced. For instant, in the case of the ${\text{I}}_{\text{Se}}$, we find that $d_{1} = 3.37$ \AA\ and $d_{2} = 3.09$ \AA, resulting in that $\Delta_{y}=0.44$ \AA. This value is larger than that of the pristine system ($\Delta_{y}=0.34$ \AA), indicating that strong enhancement of the in-plane ferroelectricity is achieved. In contrast, the formation of the $X_{\text{H}}$ results in a square piramidal-like structure [Fig. 1(d)], where the average bondlength between the Sn or Se and halogen $X$ atoms ($\bar{d}_{\text{Sn-X}}$ ,$\bar{d}_{\text{Se-X}}$) is substantial [see Table 2]. However, there is no in-plane distortion observed in the $X_{\text{H}}$, indicating that the $X_{\text{H}}$ dosn't support the in-plane ferroelectricity. Because the in-plane ferroelectricity plays an important role for inducing the PSH states, in the following discussion we will focus only for the $X_{\text{Se}}$.

Figure 3 shows the electronic band structures of the doping ($X_{\text{Se}}$) systems compared with those of the pristine one. In contrast to the pristine system [Figs. 3(a) and 3(e)], we identify impurity states in the band structures of the $X_{\text{Se}}$, which are located close to the CBM around the Fermi level [Figs. 3(b)-(d)]. This indicates that $n$-type doping is achieved, which is similar to the recent prediction of the $n$-type system on bulk SnS doped by Cl atom (experiment) \cite{Yanagi} and bulk SnSe doped by Br and I atoms (theory) \cite{YZhou}. More importantly, we find spin-split bands in the impurity states when the SOC is taken into account [Figs. 3(b)-(d)], indicating that this system is promising for spintronics. 

\begin{figure}
	\centering
		\includegraphics[width=0.6\textwidth]{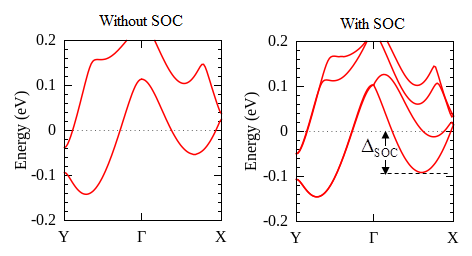}
	\caption{The band structures of the I$_{\text{Se}}$ around the Fermi level calculated without (left)and with (right) the SOC. Here, the spin splitting energy $\Delta_{\text{SOC}}$ is indicated. }
	\label{figure:Figure4}
\end{figure}

To analyse the properties of the spin-split bands in the impurity states of the $X_{\text{Se}}$, we choose the I$_{\text{Se}}$ as a representative example. Here, we focused on the spin-split impurity bands around the Fermi level shown in Fig. 4. We  find that large spin splitting is observed along the $\Gamma-X$ line, while it is extremely small (almost degenerate) along the the $\Gamma-Y$ line. Here, the spin splitting energy $\Delta_{\text{SOC}}$ of about 98.2 meV is achieved, which is much larger than that observed on the pristine system ($\Delta_{\text{SOC}}=51.6$ meV). The significanty large $\Delta_{\text{SOC}}$ found in the I$_{\text{Se}}$ is due to the increased in-plane ferroelectric distortion $\Delta_y$ [see Table 2]. Accordingly, the magnitude of the in-plane electric field $E_{y}$ increases, which is responsible for amplfying the spin splitting energy. In fact, manipulating of the spin splitting energy by the electric field has been recently reported on various 2D TMDs materials \cite{QFYao,QZhang}.  

\begin{figure}
	\centering
		\includegraphics[width=1.0\textwidth]{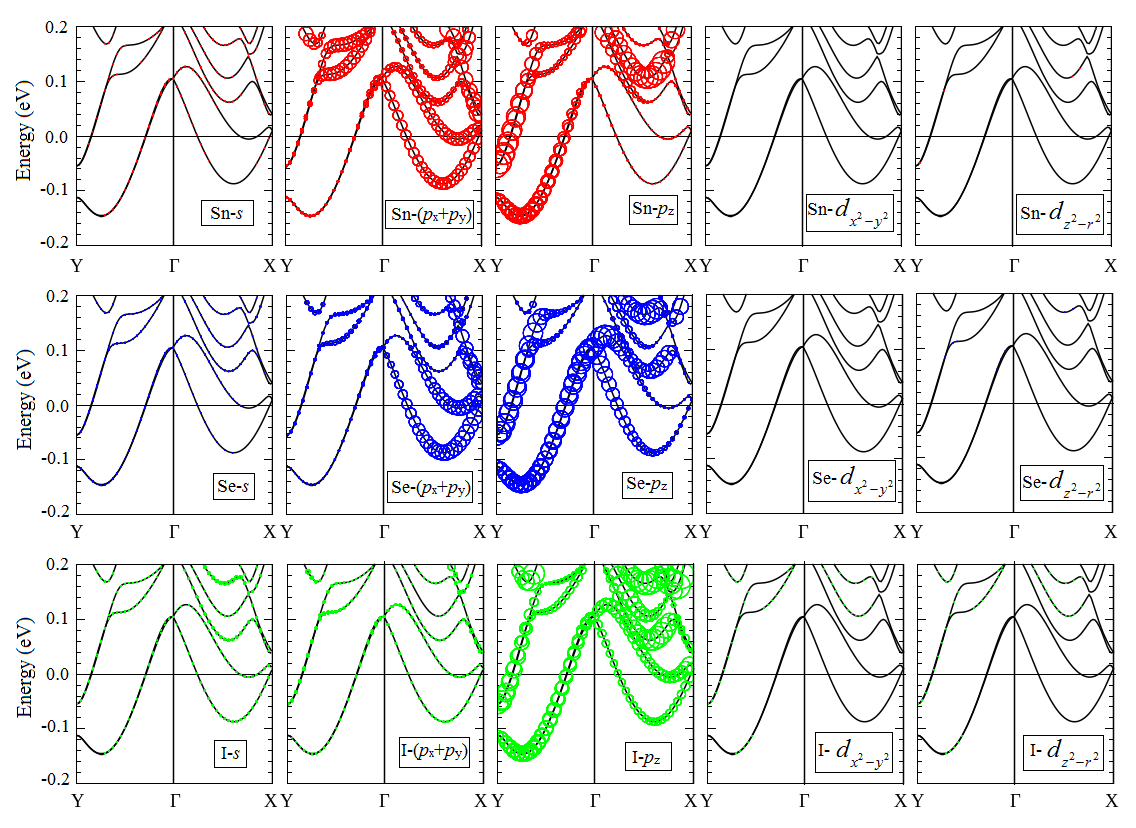}
	\caption{Orbital resolved of the spin-split impurity bands of the I$_{\text{Se}}$ projected to the atoms near impurity site. The red, blu, and green colors indicate the Sn, Se, and I atoms, respectively. The circle radius reflects the spectral weight of the specific orbital to the band.}
	\label{figure:Figure5}
\end{figure}

The microscopic origin of the spin-splitting in the impurity states can be understood by considering orbital hybridizations. Here, coupling between atomic orbitals contributes to the non-zero SOC matrix element through the relation $\zeta_{l}\left\langle \vec{L}\cdot \vec{S}\right\rangle_{u,v}$, where $\zeta_{l}$ is angular momentum resolved atomic SOC strength with $l=(s,p,d)$, $\vec{L}$ and $\vec{S}$ are the orbital angular momentum and Pauli spin operators, and $(u,v)$ is the atomic orbitals. Therefore, only the orbitals with magnetic quantum number $m_{l}\neq 0$ will contribute to the spin splitting. By calculating the orbital-resolved of the electronic band structures projected to the atom around the impurity site, we find that large spin splitting along the $\Gamma$-$X$ line is mostly originated from the contribution of the $p_{x}+p_{y}$ ($m_{l}=\pm 1$) orbitals of the Sn, Se, and I atoms [Fig. 5]. On the contrary, along the $\Gamma-Y$ line, the orbitals are dominated by the $p_{z}$ ($m_{l}=0$) orbitals, which contributes only minimally to the spin splitting. It is noted here that there is no contribution of the $d$ orbitals ($d_{x^{2}-y^{2}}$, $d_{z^{2}-r^{2}}$) in the spin-split impurity bands, which is in contrast to halogen-doped TMDs ML \cite{Guo_S,Absor3} where the $d$ orbitals plays an important role for inducing the spin-split impurity bands.  

To clearly demonstrate the properties of the spin-split impurity bands, we calculate $k$-space spin textures by using the spin density matrix of the spinor wave functions obtained from our DFT calculations. This method  has previously been successfully applied on strained WS$_{2}$ ML \cite{Absor2}, polar WSSe ML \cite{Absor5} and WSTe ML \cite{Absor4} and halogen doped PtSe$_{2}$ ML\cite{Absor3}. Here, we focus on the spin textures in the spin-split bands at Fermi level around the $\Gamma$ point [Fig. 4]. We find that the Fermi surface is characterized by the shifted two loops [Fig. 6(a)], dominated by the out-of-plane spin orientations [Figs. 6(b)-(d)]. These typical Fermi surface and spin textures indicate that the PSH states are achieved, which is similar to those observed on the (110)-oriented GaAs/AlGaAs QW \cite{Chen} and the ZnO (10-10) surface \cite{Absor1}. As mentioned previously that the I$_{\text{Se}}$ provides the in-plane electric field $E_{y}$. This leads to the fact that the SOF is enforced to be parallel with spins in the out-of-plane direction. This inhibits the precession of the spins, thereby increasing the spin relaxation time. A similar mechanism behind long spin relaxation times has been reported on [110]-oriented zinc-blende QWs \cite {Dohrmann,Couto,Ohno,Chen}, rendering that this system could provide an efficient spintronics device.    

\begin{figure}
	\centering
		\includegraphics[width=0.9\textwidth]{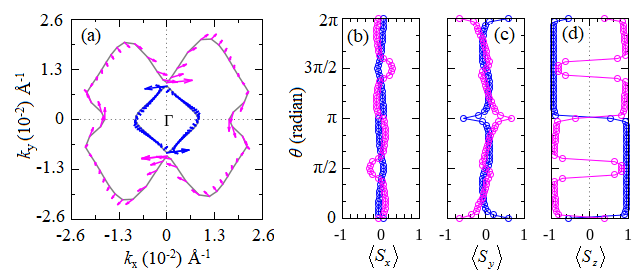}
	\caption{(a) The calculated spin textures projected to the 2D $(k_{x}-k_{y})$ plane are shown. The blue and pink arrows indicates the spin textures for the upper and lower bands, respectively. The expectation values for: (b)$\left\langle S_{x}\right\rangle$, (c) $\left\langle S_{y}\right\rangle$, and (d) $\left\langle S_{z}\right\rangle$ are given. The blue and pink colors reflect the expectation value of spin calculated for the upper and lower bands, respectively.}
	\label{figure:Figure6}
\end{figure}

\begin{table}[ht!]
\caption{Transformation rules for wave vector $\vec{k}$, and spin vector $\vec{\sigma}$ under mirror symmetry $M_{yz}$ and time reversal symmetry $T$ operations. } 
\centering 
\begin{tabular}{c c c} 
\hline\hline 
  Symmetry operation & ($k_{x}$,$k_{y}$,$k_{z}$)   & ($\sigma_{x}$,$\sigma_{y}$,$\sigma_{z}$)  \\ 
\hline 
$\hat{T}=i\sigma_{y}K$   &  ($-k_{x}$,$-k_{y}$,$-k_{z}$)  &  ($-\sigma_{x}$,$-\sigma_{y}$,$-\sigma_{z}$)  \\                                                       
$\hat{M}_{yz}=i\sigma_{x}$  &   ($-k_{x}$,$k_{y}$,$k_{z}$)  & ($\sigma_{x}$,$-\sigma_{y}$,$-\sigma_{z}$) \\ 
\hline\hline 
\end{tabular}
\label{table:Table 1} 
\end{table}

The observed spin splitting and spin textures can be explained in term of an effective low-energy  Hamiltonian, which can be deduced by considering the crystal symmetry. As mentioned previously that the relaxed $X_{\text{Se}}$ comprises only a mirror symmetry $M_{yz}$ [Fig. 1(c)]. Therefore, around the $\Gamma$ point, the wave vectors $\vec{k}$ and spin vectors $\vec{\sigma}$ can be transformed according to the miror symmetry operator $\hat{M}_{yz}=i\sigma_{x}$ and time reversal symmetry operator $\hat{T}=i\sigma_{y}K$, where $K$ is the is complex conjugation. The corresponding transformations for $\vec{k}$ and $\vec{\sigma}$ are given in Table III.  Collecting all the terms which are invariant with respect to the symmetry operations we obtain the effective Hamiltonian as follows \cite{Absor1}: 
\begin{equation}
\label{6}
H_{\Gamma}(k)=E_{0}(k)+ \alpha k_{x}\sigma_{y}+\beta k_{y}\sigma_{x}+ \gamma k_{x}\sigma_{z},
\end{equation}
where $\alpha$,$\beta$, and $\gamma$ are three independent coefficients and $E_{0}(k)=\hbar^{2}({{k_{x}^{2}}/m_{x}^{*}}+{{k_{y}^{2}}/m_{y}^{*}})/2$ is the nearly free-electron energy. The Eq.(\ref{6}) shows a combination between Rashba and Dresselhauss SOC with the coupling constants are given by $\alpha_{\text{R}}=(\beta-\alpha)/2$ and $\beta_{\text{D}}=(\beta+\alpha)/2$ parameters, respectively, and one additional anisotropic  (which is also referred as the anisotropic Drreselhauss term) parameter $\gamma$. Solving the eigenvalue problem involving the Hamiltonian of the Eq.(\ref{6}), we find that 
\begin{equation}
\label{7}
E_{\pm}(k)=E_{0}(k) \pm \sqrt{(\beta_{\text{D}}^{2}+\gamma^{2})k_{x}^{2}+\alpha_{\text{R}}^{2}k_{y}^{2}}.
\end{equation} 
Morover, the spin textures can be deduced from the averaged components of the spinor operators $\left\langle\vec{S}\right\rangle\ $, which is found to be
\begin{equation}
\label{8} 
 \left(\begin{array}{r}
\left\langle S_{x}\right\rangle_{\pm}\\
\left\langle S_{y}\right\rangle_{\pm}\\
\left\langle S_{z}\right\rangle_{\pm}
\end{array}\right)=  \left(\begin{array}{r}
\pm\sin \theta \cos\zeta\\
\pm\sin \theta \sin\zeta\\
\pm\cos\theta\end{array}\right)
\end{equation} 
where $\theta$ and $\zeta$ are defined by $\tan\theta=\sqrt{(\beta_{\text{D}}^{2}k_{x}^{2}+\alpha_{\text{R}}^{2}k_{y}^{2}})/\gamma k_{x}$ and $\tan\zeta=(\beta_{\text{D}}/\alpha_{\text{R}})\cot\phi_{k}$, respectively, with $\phi_{k}$ is related to $\vec{k}=\sqrt{k_{x}^{2}+k_{y}^{2}}(\cos \phi_{k}, \sin \phi_{k})$. 

According to the Eq. (\ref{7}), the PSH states is achieved when $\alpha_{\text{R}}$ is zero. Therefore, the spin-orbit strength of the PSH $\alpha_{PSH}=\sqrt{(\beta_{\text{D}}^{2}+\gamma^{2})}$. This means that the PSH states is purely characterized by the Dresselhauss effect, which is similar to the (110)-oriented QW. As a result, strong anisotropic spin splitting is observed, i.e., the band structure has large spin splitting along the $\Gamma-X$ (i.e $k_{x}$) line but is degenerated along the $\Gamma-Y$ (i.e $k_{y}$) line, which is in fact consistent with our calculated result of spin-split bands shown in Fig. 4. By fitting the calculated band dispersion in Fig. 4 along the $\Gamma-X$ line with Eq.(\ref{7}), we find that $\alpha_{PSH}=\sqrt{(\beta_{\text{D}}^{2}+\gamma^{2})}=1.66$ eV\AA . Moreover, from Eq. (\ref{8}), we can estimate the ratio between $\beta_{\text{D}}$ and $\gamma$ by comparing the in-plane and out-of-plane spin component along the $\Gamma-X$ line, and found that $\beta_{\text{D}}/\gamma=0.036$. Therefore, by using $\sqrt{(\beta_{\text{D}}^{2}+\gamma^{2})}=1.66$, we find that $\beta_{\text{D}}=6.04$ meV\AA\ and $\gamma=1.659$ eV\AA. The larger value of $\gamma$ indicates that the spin-split states are dominated by the out-of-plane spin component, which is also agree well with the calculated spin textures shown in Figs. 6(b)-(d). 

\begin{table}[ht!]
\caption{Several selected materials and parameters characterizing the spin-orbit strength of the PSH $\alpha_{PSH}$ (in eV\AA) and the wavelength  of the PSH $\lambda_{PSH}$ (in nm)} 
\centering 
\begin{tabular}{c c c c} 
\hline\hline 
  Systems & $\alpha_{PSH}$ (eV\AA)  & $\lambda_{PSH}$ (nm)  & Reference \\ 
\hline 
SnSe-I ML   & 1.76   & 1.27 & This work  \\        
SnSe-Br ML & 1.65 & 1.35  & This work   \\
SnSe-Cl ML & 1.60 & 1.41& This work\\        	
GaAs/AlGaAs QW & (3.5-4.9)$\times 10^{-3}$  & (7.3-10) $\times10^{3}$& Ref.\cite{Walser} \\
               & 2.77 $\times10^{-3}$ & 5.5$\times10^{3}$ & Ref.\cite{Schonhuber} \\
InAlAs/InGaAs QW & 1.0 $\times10^{-3}$&      & Ref.\cite{Ishihara}\\
              & 2.0 $\times10^{-3}$&      & Ref.\cite{Sasaki}\\
CdTe QW       &    &   5.6$\times10^{3}$ & Ref.\cite{Passmaan}\\
ZnO(10-10) surface & 34.78 $\times10^{-3}$& 1.9$\times10^{2}$ & Ref.\cite{Absor1}\\
bulk BiInO$_{3}$ &1.91 & 2.0 & Ref.\cite{LLTao}\\
\hline\hline 
\end{tabular}
\label{table:Table 2} 
\end{table}

Now, we calculate another important parameter in the PSH states called as the wave length $\lambda_{PSH}$, which characterizes the spatially periodic mode \cite{Bernevig}. Here, $\lambda_{PSH}$ is defined as $\lambda_{PSH}=(\pi\hbar^{2})/2m^{*}\alpha_{PSH}$, where $m^{*}$ is the effective mass \cite{Bernevig}. By analyzing the dispersion of the spin-split bands around the Fermi level along the $\Gamma$-$X$ direction, we find that $m^{*}=0.67m_{0}$, where $m_{0}$ is the free electron mass, resulting in that $\lambda_{PSH}$ is about 1.27 nm. The calculated $\lambda_{PSH}$ is typically on the scale of the lithographic dimension used in the recent semiconductor industry \cite{Fiori}.

We summarize the calculated results of the $\alpha_{PSH}$ and $\lambda_{PSH}$ in Table IV and compare the result with a few selected PSH materials from previously reported data. This revealed that the value of $\alpha_{PSH}$ is much larger than those observed in the PSH of various zinc-blende $n$-type QW structures such as GaAs/AlGaAs \cite{Walser,Schonhuber} and InAlAs/InGaAs \cite{Sasaki,Ishihara}, ZnO (10-10) surface\cite{Absor1}, strained LaAlO3/SrTiO3 (001) interface \cite{Yamaguchi}, and is comparable with the recently reported PSH states on BiInO$_{3}$ \cite{LLTao}. The large value of $\alpha_{PSH}$ should ensure that a small $\lambda_{PSH}$ is achieved. Indeed, the calculated value of $\lambda_{PSH}$ is in fact three order less than that observed on the GaAs/AlGaAs QWs \cite{Walser,Schonhuber} and CdTe QW\cite{Passmaan}. Moreover, the calculated $\lambda_{PSH}$ is much smaller than the electron coherence length (0.1$\mu$m) measured at room temperature \cite{Wide}, which is important for the development of a high-density scalable spintronics devices operating at room temperatures. 

Thus far, we have found that the PSH with large spin splitting is achieved in the SnSe ML functionalized by a substitutional halogen doping. Because the PSH states is achieved on the spin-split bands near Fermi level [Fig. 3(c)], $n$-type doping for spintronics is expected to be realized. Therefore, it enables us to allow operation as a spin-field effect transistor device at room temperature \cite {Yaji}. It is noted here that our method for inducing the PSH states with large spin splitting can be extendable to other 2D $MX$ monochalcogenide materials such as SnS, GeS, and GeSe where the electronic properties are similar \cite {Gomes1}. Therefore, this work provides a possible way to induce the PSH states with large spin splitting in the 2D nanomaterials, which is promising for future spintronic applications.

\section{CONCLUSION}
In summary, we have investigated the effect of a substitutional halogen doping on the electronic properties of the SnSe ML by employing the first-principles DFT calculations. We found that the substitutional halogen impurity can induce the PSH states with very large spin splitting, where $k$-space Fermi surface is characterized by the shifted two loops exhibiting the out-of-plane spin orientations. We showed that these PSH states is originated from in-plane electric field induced by in-plane ferroelectric distortion. We clarified the properties of the PSH states in term of an effective $\vec{k}\cdot \vec{p}$ Hamiltonian obtained from symmetry consideration. Recently, the doped group IV-monochalcogenides ML has been extensively studied \cite{Chao,WangQi}. Our study clarified that the halogen doping plays an important role for inducing the PSH states in the electronic properties of the SnSe ML, which is expected to be useful for the development of an efficient and a high-density scalable spintronic devices operating at room temperatures. Finally, we expect that our theoretical predictions will stimulate experimental efforts in the exploration of the PSH in the 2D materials, which functional properties may be useful for future spintronic applications.

\begin{acknowledgments}

The first authors (M. A. U. Absor) would like to thank T. Ozaki for his permission to use the computer facilities in the ISSP, the University of Tokyo, Japan, as a part of the ISS2018 summer school project. This work was partly supported by PDUPT 2018 Research Grant funded by the Ministry of Research and Technology and Higher Education (RISTEK-DIKTI), Republic of Indonesia. 

\end{acknowledgments}

\bibliography{Reference1}


\end{document}